# Structural, Vibrational, Elastic and Topological Properties of *PaN* Under Pressure

P. Modak[1], Ashok K. Verma[1], A. Svane[2], N. E. Christensen[2] and Surinder M. Sharma[1]

[1]High Pressure and Synchrotron Radiation Physics Division, Bhabha Atomic Research Centre, Trombay, Mumbai-400085, India

[2]Dept. of Physics and Astronomy, University of Aarhus, DK 8000, Aarhus C, Denmark

**Abstract**

Electronic, structural, vibrational and elastic properties of *PaN* have been studied both at ambient and high pressures, using first principles methods with several commonly used parameterizations of the exchange-correlation energy. The generalized gradient approximation (*GGA*) reproduces the ground state properties satisfactorily. Under pressure *PaN* is found to undergo a structural transition from *NaCl* to the *R-3m* structure near *58 GPa*. The high pressure behavior of the acoustic phonon branch along the (1,0,0) and (1,1,0) directions, and the $C_{44}$ elastic constant are anomalous, which signals the structural transition. With *GGA* exchange-correlation, a topological transition in the charge density occurs near the structural transition which may be regarded as a quantum phase transition, where the order parameter obeys a mean field scaling law. However, the topological transition is absent when other exchange-correlation functionals are invoked (local density approximation (*LDA*) and hybrid functional). Therefore, this constitutes an example of *GGA* and *LDA* leading to qualitatively different predictions, and it is of great interest to examine experimentally whether this topological transition occurs.

I. INTRODUCTION

In recent times actinide nitrides have received considerable attention because of their possible use as advanced nuclear fuel materials.[1] There are, however, very few studies of the properties of the mono-nitride of protactinium, *PaN*.[2-4] *Pa* is one of the rarest and most expensive naturally occurring elements. It is highly radioactive and toxic, which may be the reason for the limited number of studies available for this element and its compounds.[2-7] Theoretically, it is a challenge to treat 5*f* electrons within density functional theory (*DFT*[8]) as their behavior is intermediate between itinerant, like that in transition metals, and localized, like that in rare earths. Recent developments of approaches like the inclusion of self-interaction corrections (*SIC*)[9] or on-site Hubbard type interactions in standard *DFT*, the *GW*,[10] and the dynamical mean field theory (*DMFT*)[11] aim to overcome the shortcomings of standard *DFT,* and have spawned renewed interest in 5*f* elements and their compounds. It is known that in light actinides (*Th* to *Np*) the 5*f* electrons are itinerant, whereas in the heavier actinides (*Am* and beyond) the 5*f* electrons are localized.[3] Protactinium is the first element of the actinide series where the 5*f* occupation exceeds one, and earlier studies[6,7] revealed that under pressure the *f* occupancy of *Pa* metal increases causing a sequence of structural transitions in the sequence $bct \rightarrow \alpha\text{-}U \rightarrow bct \rightarrow hcp$. Therefore it will be interesting to study the effect of pressure on *f* occupancy also in *Pa* compounds, and its subsequent effects on their structural stability. This is the issue addressed in the present work in the case of *PaN*.

At ambient conditions all the actinide mono-nitrides crystallize in the *NaCl* (*B1*) structure[3] irrespective of the nature of the 5*f* electrons of the actinide metals. *PaN* is the first member of the actinide mono-nitride series where the 5*f* electrons actively participate in the chemical bonding, which leads to the anomalously short lattice constant compared to the heavier actinide mono-nitrides[3]. As the 5f electrons in *PaN* are itinerant the *DFT* is expected to provide a valid description of their bonding properties and lead to accurate ground state and high pressure properties. However, it is well known that the local density approximation (*LDA*)[12] to *DFT* is poor for *f* electron systems which often have large anisotropic charge distributions. Therefore a comparative study of the applicability of *LDA* versus the generalized gradient approximation (*GGA*)[13] for the exchange-correlation effects, as well as recent approaches like Hubbard *U* corrected funtionals (*LDA+U, GGA+U*) and hybrid functionals will be interesting. In our earlier study[14] on *ThN* and *UN* we showed that the *GGA* is more accurate than the *LDA* for both compounds although the differences in the predicted equilibrium properties are smaller for *ThN*. Also, *LDA* underestimates the phase transition pressures in both cases, but the pressure induced structural sequences are correctly reproduced. Furthermore, a distinctly different high pressure behavior was found for these two compounds, primarily due to the active participation of the 5*f* electrons in the chemical bonding in *UN*.[14] *PaN* has an *f* occupancy between those of *ThN* and *UN*, and it is interesting to examine to what extent the behavior of *PaN* under pressure follows that of *UN* or that of *ThN*. The first member of the 4*f* mono-nitride series, *CeN*, has identical *f* occupancy and an anomalously short lattice constant, similar to *PaN*, yet is observed[15] to undergo a *B1-B2* structural transition at high pressure similar to that observed in *ThN*.[14] Therefore, high pressure studies of *PaN* have great importance in order to understand the role of the *f* electrons in the high pressure structural stability of *f*-electron systems. Further, it may also shed light on the role of the 6*d* electrons in pressure induced structural phase transitions, since *Ce* and *Pa* are iso-electronic except for the single extra 6*d* electron in *Pa*. Also the study of the high pressure behavior of elastic and vibrational properties of this system will be important in order to identify the origin of the pressure induced structural transitions.

In this work the structural properties of *PaN* under pressure have been studied by calculation of enthalpies using both *LDA* and *GGA* exchange-correlation functionals. It is found that at high pressure *PaN* follows *UN* and undergoes a structural transition from the *B1* to the *R-3m* structure. The *GGA* transition pressure is *58 GPa*. The high pressure behavior of the calculated acoustic phonon dispersion and the $C_{44}$ elastic constant are anomalous, and the $C_{44}$ elastic constant is softening few *GPa* above the pressure of the structural transformation. The electronic charge density is a quantum observable, which is supposed to be calculable from *DFT*, and which may also be determined experimentally via x-ray or combined x-ray and polarized neutron diffraction techniques for magnetic systems.[16,17] Analyzing the calculated charge density of *PaN*, we have found that for the *GGA* functional a topological transition in charge density occurs close to the structural transformation. This is similar to a high-pressure iso-structural transition in *Ca*.[18] For *PaN*, the structural transition occurs for all parameterizations of *DFT* considered here, i.e *LDA*, *GGA*, *LDA/GGA+U*, *LDA/GGA+SO+U* (*SO*: spin orbit interaction) and the hybrid method. In contrast, the topological transition in charge density is obtained only for *GGA* based approaches i.e. *GGA*, *GGA+U* and *GGA+SO+U*, and it remains a question whether it is an artifact of the *GGA* approaches, or a genuine physical effect, which is observable but captured neither by the *LDA* nor by the hybrid functional.

## II. METHOD OF CALCULATIONS

To determine the high pressure structural stability of the *B1*, *B2* and *R-3m* structures for *PaN* the enthalpies were calculated at different pressure and as a function of the rhombohedral angle, using the Vienna *Ab-initio* Simulation Package (*VASP*).[19-22] A rhombohedral cell was considered because all three structures studied may be cast into this cell by varying the cell angle. Phonon dispersion relations were calculated within the small-displacement force method using a *3×3×3* supercell. The displacement pattern was determined and the dynamical matrix solved using the *PHON* software.[23] The elastic constants were determined from energy-strain relations which were obtained by calculation of the total energy for different strains, like volume, tetragonal and trigonal strains. The details of the calculations of phonon dispersions and elastic constants can be found in our earlier study[14] on *ThN* and *UN*. For all these calculations the projector augmented wave (*PAW*) potentials with *1000 eV* plane-wave energy cut-off were used. In the *PAW* potentials $5f^2$, $6p^6$, $6d^1$, $7s^2$; and $2s^2$, $2p^3$ were considered as the valence configurations for the *Pa*- and *N*-atom, respectively. The Brillouin zone (*BZ*) integration was carried out using a uniform Monkhorst-Pack[24] ***k***-point grid (*20×20×20* for the *NaCl* structure). The *GGA-PBE*[13] parameterization of the exchange-correlation functional was used for the majority of calculations, while we also checked the effect of on-site Coulomb correlations for the *f* electrons and the spin-orbit (*SO*) interaction on the equilibrium properties and on the phase stability. The *Critic* program[25] was used to analyze the charge density topology. Some of the calculations were repeated with *LDA*, *LDA+U*, *LDA+SO+U* and hybrid functional approaches to identify the presence or absence of the structural and topological transitions in *PaN* under pressure. In the hybrid functional approach a certain fraction of the approximate *DFT* exchange energy is replaced by the exact Hartree-Fock exchange energy.[26] Furthermore, some of these calculations were checked, in particular the charge density analysis, with an all electron method, the full potential linear augmented plane wave method as implemented in the *WIEN2k*[27] program. To obtain an accurate charge density in these calculations the spherical harmonics expansion was increased to $l_{max}=10$.

## III. RESULTS

The calculated equilibrium lattice constant of *PaN* (*4.97 Å*) and bulk modulus (*224.8 GPa*) with the *GGA* exchange-correlation functional are in good agreement with experiment[2] and previous calculations[4] (*4.96 Å* and *223 GPa,* respectively). The experimental lattice constant[2] is *5.05 Å,* while bulk modulus data is not available. The inclusion of on-site Hubbard type correlations for the 5*f* states (i.e., *GGA+U* with $U_{eff}(5f)=U-J=4$ eV) gives a very accurate equilibrium lattice constant (*5.04 Å*). Since the *GGA* underestimation of the lattice constant is small and the main focus of the present work is on the high pressure behavior *PaN*, the Hubbard interactions are not included in the calculations to follow unless specified, as this effect is expected to decrease under pressure. The *SO* effects on equilibrium properties are small. The *LDA* lattice constant (*4.88 Å*) underestimates the experimental value significantly (*3.4%*), and the *LDA* bulk modulus (*253 GPa*) is *12.5%* larger than that of *GGA*.

Figure1 shows the enthalpy variation with the rhombohedral angle in the *R-3m* structure at a few selected pressures. At ambient pressure the enthalpy minimum at $\alpha=60°$ corresponds to the *NaCl* structure. The *B2* structure which corresponds to $\alpha=90°$ also in a local minimum (not shown), however its enthalpy is much higher compared to that of *NaCl*. Under pressure a new local minimum develops at lower angle which becomes the absolute minimum at the transition pressure. The calculated transition pressure is *58 GPa*. The change in rhombohedral angle is finite at the transition, i.e. it is a first-order transition, however with such a small energy barrier that it probably will look like a continuous transition in experiment. Therefore a lattice distortion characterizes the transition where the atoms occupy the same crystallographic positions ((1a) and (1b) of the *R-3m* structure) before and after the transition. Just above the transition pressure the minimum corresponding to the *NaCl* structure remains a local minimum, until at a slightly higher pressure it exhibits a negative curvature with respect to variation of α, as illustrated in figure 1 (inset) for *p=65* GPa. Also we found that the *R-3m* structure is unstable at ambient pressure, *i. e.*, there is no local minimum corresponding to *R-3m*, as it converts to the *NaCl* structure under relaxation (here rhombohedral angle). We have calculated the equations of states for the *B1*, *B2* and *R-3m* structures up to *150 GPa* and found that for pressures above *58 GPa* the most stable structure occurs for an angle α≠60°, i.e. the structure has no higher symmetry than *R-3m*. At 100 *GPa* the enthalpy of *B2* structure is 0.65 eV higher than that of *R-3m* structure and the enthalpy difference decreases under pressure. Therefore there is a possibility of *B2* phase at pressure higher than 150 *GPa*. With *LDA* the transition pressure is calculated to be *50 GPa*. The structural transition occurs for all the *DFT* approaches applied, however with varying transition pressure. We hold the *GGA* results more reliable based on the accurate results obtained for the similar compounds *ThN* and *UN*.[14] This high pressure structural behavior of *PaN* is similar to that of *UN* though the transition pressure is higher in *PaN*.

Figure 2 shows the calculated phonon dispersions for *PaN* in the *NaCl* structure, both at ambient and at high pressures. It is seen that the acoustic phonons have anomalous softening behavior (frequencies decreasing with pressure) at high pressures along the *Γ—X* symmetry line of the *BZ*, and eventually turn imaginary corresponding to dynamical instability near *95 GPa*. In contrast, the optical modes show normal pressure hardening. The acoustic phonon anomaly reflects the anharmonic energy landscape depicted in Figure 1 and signals the structural instability. Similar high pressure behavior of acoustic phonons was also obtained in our earlier study[28] on Vanadium which undergoes a structural transition (*BCC* to *R-3m*) near *60 GPa*.

In a cubic crystal system, there are three independent elastic constants $C_{11}$, $C_{12}$ and $C_{44}$ which describe their elastic properties. Figure 3 shows the pressure variation of elastic constants. The $C_{11}$ and $C_{12}$ show the normal high pressure behavior, i.e., hardening with pressure whereas $C_{44}$ softens under pressure becoming zero few *GPa* above the transition pressure. In a central force approximation $C_{12}=C_{44}$, and hence any difference between these two elastic constants is due to the presence of non-central forces or many body interactions. In many body interactions the dominant contributions come from the second neighbor interactions i.e. in this case *Pa-Pa* or *N-N* repulsions. The $C_{44}$ elastic constant is related to a trigonal deformation of the cubic cell and its softening just above the transition implies that the restoring force decreases under pressure and before it becomes zero the system makes the transition, as illustrated in Figure 1. Therefore the $C_{44}$ elastic constant softening marks the mechanical instability of the lattice and signals the structural transition. In this respect the high pressure behavior of vibrational and the elastic properties of *PaN* is similar to that of *UN*.[14]

The electronic structure of *PaN* is illustrated in Figure 4. At ambient pressure, Figure 4a, the *N p* valence bands are seen in the region between -6 to -2 *eV* (relative to the Fermi level, $E_F$), while the *Pa f*-bands occur 0-2 *eV* above the Fermi level, with significant hybridization with *Pa* $d$ leading to a tail stretching below $E_F$. Two $d$- and $f$-hybridized bands cross the Fermi level and hybridize with $N\ p$ along the $\Gamma$—$X$ and $X$—$W$ symmetry lines of the *BZ*. The *PaN* bandstructure is very similar to that of *CeN* (Figure 4b), except that the *4f*-bands of *CeN* are narrower than the *5f* bands of *PaN*. Because of the different valence electron number in these compounds the Fermi level cuts a bit more into the *fd*-bands in *PaN*. The bands (doubly degenerate at *W* point) just below the Fermi level along *W-L* direction in *PaN* have large dispersions compared to the equivalent bands in *CeN* which is 0.*2 eV* above $E_F$. Also in *PaN* they are *df*-hybridized whereas in *CeN* they are pure *f* bands. At high pressure (Figure 4c) the bands generally widen, but the band topology remains the same. The two bands crossing $E_F$ along the *W-L* direction merge due to accidental degeneracy at the *L* point, and a band flattening occurs just above $E_F$ along the $\Gamma$—*L* direction. At the $\Gamma$ point the lowest unoccupied state is triply degenerate with dominantly *d* character. The degeneracy is lifted under rhombohedral distortion causing a singlet to move below $E_F$ and a doublet to move farther upwards from $E_F$ (Figure 4d). Therefore, the *Pa-d* electrons play a crucial role in this structural phase transition. The total and partial density of states (*DOS*) at $E_F$ in *NaCl* structure decreases under pressure due to band broadening. However, in the *R-3m* structure the partial *f-DOS* at $E_F$ slightly increases causing an overall increase of the total *DOS* at $E_F$ compared to that in the *NaCl* structure. Therefore a band Jahn-Teller mechanism[29] which is responsible for *B1* to *R-3m* transition in *UN*[14] may be ruled out for *PaN*. In the Jahn-Teller picture the system may lower its energy through a lattice distortion which causes symmetry related splitting of degenerate narrow bands close to the Fermi level. The energy required to distort the lattice is obtained from the gain in band energy resulting from the reduction of the *DOS* at the Fermi level which is not happening here. The total DOS at $E_F$ for *B2* structure calculated at volumes corresponding to ambient and transition pressure are much higher compared to that for *NaCl* and *R-3m* structures, which may be the reason for not favoring this structure.

To study the pressure variation of *f*-occupancy in *PaN* we have integrated the partial *DOS*s which show that the *d*-electron population in the *NaCl* structure is slightly increase under pressure although the *f*-occupation remains constant. However it is to be noted that the partial *DOS* splitting is possible only inside the atomic sphere and hence it may not be very accurate. To get an accurate estimate we have calculated the Bader charges for *Pa* and *N* atoms in *PaN* under pressure. We have found there is more and more charge transfer from *Pa* to *N* atom under pressure and close to structural transition there is 0.05*e* more charge transfer from *Pa* to *N* atom compared to that at ambient pressure. However in *R-3m* structure the pressure variation of Bader charges show opposite behavior *i.e.* decrease of charge transfer from *Pa* to *N* atom under pressure.

The charge density topology of *PaN* in the *NaCl* structure was analyzed both at ambient and at high pressure. According to the Morse topology theory[30] the electron charge density in a crystal is a three dimensional scalar quantity which can have four non-degenerate types of critical points (*CPs*), namely: local minima, local maxima and two kinds of saddle points. The *CPs* are characterized by the vanishing gradient of the electron density and are classified according to their rank and signature, *(r,s)*. The rank is defined as the number of non-zero eigenvalues of the Hessian matrix of the charge density and the signature is the algebraic sum of the signs of the eigenvalues. The sign of the eigenvalue essentially gives the curvature of the charge density iso-surface. In *3D* solids a charge density maximum is classified as a *(3,-3) CP* and often termed a nuclear *CP*, as often charge density maxima are found at atomic sites. A saddle point having one positive curvature is classified as a (*3, -1*) *CP* or bond *CP,* as it always connects two maxima. A saddle point which has two positive curvatures is classified as a (*3, 1*) *CP* or ring *CP* as it generally appears inside a ring of bonds. A minimum in the charge density topology is classified as a (*3, 3*) *CP* or cage *CP* as it represents an interior point of a cage formed by non-coplanar rings.

At ambient pressure and with *GGA* exchange-correlations the electron density topology of *PaN* was found to possess six in-equivalent critical points as shown in Figure 5. There are two nuclear *CPs* corresponding to the positions of *Pa* and *N* atoms (*Wyckoff* position *4a* and *4b* respectively), one bond *CP* at *24e*, one ring *CP* at *48g* and two cage *CPs* at *8c* and *24d*. Under pressure the ring *CPs* move towards the cage *CP* at *24d* Wyckoff position and finally coalesce with this *CP* at a pressure of *58 GPa*. This constitutes a topological transition in the charge density which may have impact on the structural instability in the system. A similar topological transition was shown to be responsible for the iso-structural transition in *Ca* at high pressure.[18] However, calculations with *LDA* and hybrid functionals fail to show any topological transition in the charge density in *PaN*. With these exchange-correlation approximations the ambient pressure charge density topology is identical to that obtained with *GGA* at high pressure. To check whether the topological transition is an artifact of the *PBE-GGA* we have calculated the charge density topology with the Wu and Cohen[31] flavor of *GGA*, and obtained the same result, i.e. also a topological transition in this case. Inclusion of Hubbard interaction and *SO* corrections with *LDA/GGA* do not change the respective *LDA/GGA* charge density topologies. Hence, the topological transition is obtained primarily due to the presence of charge density gradient related terms in the effective potential in *GGA,* which appear neither in *LDA* nor in hybrid approaches. All approaches are of course approximations to the exact *DFT*, and it remains a question whether the topological transition in *PaN* is a real effect. However, it constitutes a distinct example where *GGA* and *LDA* lead to qualitatively different results for the ground state electron density of a solid system. Eventually, the truth about the topological transition in *PaN* can be unveiled only through experiment. In contrast to these conclusions for *PaN*, the concurrent iso-structural and charge topology transitions in *Ca*,[18] have been found by the present authors to occur also in conjunction in *LDA* as well as in *GGA*.

Figure 6 presents a charge density iso-surface obtained with *LDA* and *GGA* exchange-correlations at the *GGA* equilibrium volume with the iso-value corresponding to the charge density at the ring *CP* (cyan points in Figures 5 and 8). It is seen that the iso-surfaces at the octahedral hole sites in the *LDA* case are connected by thin cylindrical surfaces, whereas they are disconnected in the *GGA* case, and a detached conical surface is present in between them. The presence of this conical surface is a result of the ring *CP* at *48g*. Figures 7a and b also show the iso-surfaces of the charge density differences, as found with *GGA* and *LDA*. It is seen that *GGA* generally leads to larger densities than *LDA* close to the atoms, where the charge density is high, whereas *LDA* leads to larger densities in the interstitial regions, where the charge density is lower, including the region where the ring and cage *CPs* occur.

As the vibrational and elastic properties of crystal are directly related to its charge density distribution, the pressure induced topological transition may be related to the anomalous phonon and elastic softening. Both metallic and ionic interactions are present in this system. Figure 8 shows the bifurcated *Pa-Pa* interaction path at ambient pressure for metallic interactions. Similar interaction paths also exist for *N-N* interactions but since the ionic radius of *Pa* is *20%* larger than that of *N*, *Pa-Pa* electrostatic interactions will be dominant or cation-cation contact would build up earlier. The presence of bifurcated interaction paths (through ring *CP*) partially screens the *Pa-Pa* metallic interactions causing smaller restoring force for trigonal deformation. With increasing pressure the movement of this *CP* towards the cage *CP* (at the *24d Wyckoff* positions) increases the angle of the interaction path causing increased screening and finally, when the critical pressure is reached, the ring *CP* and the cage *CP* coalesce and a straight interaction path with maximum screening is formed. However, the decrease of the *Pa-Pa* distance under pressure will give increased electrostatic repulsion which also may favor trigonal distortion. Hence the role of the charge density topological transition on the high pressure phonon anomaly and the elastic softening is not clear. Since other exchange-correlations (*LDA* and hybrid functional) lead to the structural transition but do not find the topological transition, it is concluded that the topological transition is not responsible for the structural transition.

The pressure induced topological transition in the electronic charge density can be regarded as a quantum phase transition. In this case the distance ($l$) between the ring *CP* and the cage *CP* plays the role of the order parameter, which smoothly goes to zero (figure 9) as the controlling parameter, here pressure, reaches the critical value at the quantum critical point (*QCP*). The charge density was calculated for different volumes very close to the topological transition and both $l$ and the reduced pressure ($p=|P-P_c|/P_c$) were determined. It was found that $l$ and $p$ obey the scaling law $l \sim p^{1/\delta}$, where the critical exponent $\delta$ in mean field theory has the value *3* for a *3D* scalar field. Fitting to this scaling law the value of *1/δ~0.37* was found for a reduced pressure of $p=3\times10^{-6}$, i.e. *δ* is very close to *3*.

IV. CONCLUSIONS

The high pressure behavior of *PaN* has been studied using first principles electronic structure methods and a structural transition from *B1* to the *R-3m* structure was predicted at a pressure of *58 GPa*. The high pressure behavior of the acoustic phonons is anomalous and a softening of the $C_{44}$ elastic constant under pressure is responsible for the mechanical instability of the *B1* phase and the corresponding phase transition to *R-3m* structure. From analysis of the electronic structure the *Pa-d* electrons are found to play a crucial role for the structural transition, as *d* related bands close to $E_F$ are changing under pressure and $d\text{-}t_{2g}$ state splits due to the phase transition. A band Jahn-Teller mechanism cannot be responsible for the structural transition as the *DOS* at $E_F$ increases in the *R-3m* structure. A topological analysis of the electron density showed that near the structural transition (*58 GPa)* there is a topological transition in the charge density which may be regarded as a quantum phase transition within the *B1* structure. An order parameter for the topological transition has been identified, which obeys a scaling law with a mean field exponent. The topological transition occurs for two versions of *GGA*, but is absent for the *LDA* and hybrid approximations to the exchange-correlations functional, while the structural transition is seen with all functionals, albeit at varying transition pressure. This indicates that the topological transition is not related to the structural transition. Further experiments are needed to show whether the topological transition in *PaN* is a genuine physical effect or occurs as an artifact of the *GGA* functional, which contains potential terms with explicit dependence on the gradient of the charge density.

**REFERENCES**

1. Y. Arai and K. Minato, J. Nucl. Mater. **344**, 180 (2005).

2. J. Bohet and W. Müller, J. Less-Common Met. **57**, 185 (1978).

3. M. S. S. Brooks, J. Phys. F: Met. Phys., **14**, 857 (1984).

4. R. Atta-Fynn and A. K. Ray, Phys. Rev. B **76**, 115101 (2007).

5. K. O. Obodo and N. Chetty, J. Phys.: Condens. Matter **25**, 145603(2013)

6. P. Söderlind and O. Eriksson, Phys. Rev. B**56**, 10719 (1997).

7. R. G. Haire, S. Heathman, M. Idiri, T. Le Bihan, A. Lindbaum and J. Rebizant, AIP Conf. Proc. **673**, 85 (2003); Phys. Rev. B **67**, 134101 (2003).

8. W. Kohn and L. J. Sham, Phys. Rev. **140**, A1133 (1965).

9. L. Petit, A. Svane, Z. Szotek, W. M. Temmerman and G. M. Stocks, Phys. Rev. B **80**, 045124 (2009).

10. A. Svane, R. C. Albers, N. E. Christensen, M. van Schilfgaarde, A. N. Chantis and J.-X. Zhu, Phys. Rev. B **87**, 045109 (2013).

11. S. Y. Savrasov, K. Haule and G. Kotliar, Phys. Rev. Lett. **96**, 036404 (2006).

12. J.P. Perdew, and Y. Wang, Phys. Rev. B **45**, 13244 (1992).

13. J. P. Perdew, K. Burke and M. Ernzerhof, Phys. Rev. Lett. **77**, 3865 (1996).

14. P. Modak and A. K. Verma, Phys. Rev. B **84**, 024108 (2011).

15. J. Staun Olsen, J. –E. Jørgensen, L. Gerward, G. Vaitheeswaran, V. Kanchana and A. Svane, J. Alloys and Comp. **533**, 29 (2012).

16. L. J. Farrugia and C. Evans, J. Phys. Chem. A**109**, 8834 (2005).

17. S. Mondal, S. van Smaalen, A. Schönleber, Y. Filinchuk, D. Chernyshov, S. I. Simak, A. S. Mikhaylushkin, I. A. Abrikosov, E. Zarechnaya, L. Dubrovinsky and N. Dubrovinskaia, Phys. Rev. Lett. **106**, 215502 (2011).

18. T. E. Jones, M. E. Eberhart, and D. P. Clougherty, Phys. Rev. Lett. **105**, 265702 (2010).

19. G. Kresse and J. Hafner, J. Phys.: Condens. Matter **6**, 8245 (1994).

20. G. Kresse and J. Furthmüller, Comput. Mater. Sci. **6**, 15 (1996).

21. P. E. Blöchl, Phys. Rev. B **50**, 17953 (1994).

22. G. Kresse, and D. Joubert, Phys. Rev. B**59**, 1758 (1999).

23. D. Alfè, (1998). Program available at http://chianti.geol.ucl.ac.uk/~dario.

24. H. J. Monkhorst and J. D. Pack, Phys. Rev. B **13**, 5188 (1976).

25. A. Otero de-la-Roza, M. A. Blanco, A. M. Pendás and V. Luaña, Comput. Phys. Commun. **180**, 157 (2009).

26. A. D. Becke, J. Chem. Phys. **98**, 1372 (1993).

27. P. Blaha, K. Schwarz, G. K. H. Madsen, D. Kvasnicka and J. Luitz, *WIEN2K, An Augmented Plane Wave + Local Orbitals Program for Calculating Crystal Properties,* edited by K. Schwarz ( TU Wien, Austria 2001).

28. A. K. Verma and P. Modak, Europhys. Lett. **81**, 37003 (2008).

29. I. B.Bersuker, *TheJahn-Teller Effect* (CambridgeUniversity Press) 2006.

30. J. Milnor, *Morse Theory* (Princeton University Press, ISBN 0-69-08008-9) 1963.

31. Z. Wu and R. E. Cohen, Phys. Rev. B **73**, 235116 (2006).

**Figure Captions**

**Figure 1**: Enthalpy variation of *PaN* (relative to the *NaCl* structure) as a function of the rhombohedral angle α at several pressures. The inset shows a blow up around α=*60°* to highlight the curvature change around this point under pressure. The number on each curve represents the pressure in *GPa* (color online).

**Figure 2**: Phonon dispersions for *PaN* in the *B1* structure along high symmetry directions of the *BZ* (color online).

**Figure 3**: Pressure variation of the elastic constants of *PaN* in the *B1* structure (color online).

**Figure 4**: Electronic band structures along some high symmetry directions of the *BZ* and density of states (states/eV/cell). (a): for *PaN* at ambient pressure in the *B1* structure; (b): for *CeN* at ambient pressure in the *B1* structure; (c): for *PaN* at high pressure (*P=70 GPa*) in the *B1* structure; (d): for *PaN* at high pressure (*P=70 GPa*) in the *R-3m* structure (color online).

**Figure 5**: Critical points (*CP*) of *PaN* in the *NaCl* structure at ambient pressure. Green (N1) and violet (N2) spheres are nuclear *CPs* corresponding to *Pa* and *N* atoms respectively. Red (B) spheres are bond *CPs*, cyan (R) spheres are ring *CPs*, yellow (C1) and pink (C2) spheres are two non-equivalent cage *CPs* (color online).

**Figure 6**: *3D* charge density plot of *PaN* in the *B1* structure for $V=30.75 \text{ Å}^3$, using *GGA* (a) and *LDA* (b) at the iso-value $\rho=0.1484e$. Large and small spheres are *Pa* and *N* atoms, respectively (color online).

**Figure 7**: *3D* charge density difference plot ($\rho^{GGA} - \rho^{LDA}$) for iso-value *0.004e* (a), and for iso-value *-0.004e* (b). Two sides of each surface have different color. Large and small spheres are *Pa* and *N* atoms, respectively (color online).

**Figure 8**: *Pa-Pa* interaction paths. Green (N1) and violet (N2) spheres are nuclear *CPs* corresponding to *Pa* and *N* respectively. Cyan (R) spheres are ring *CPs* and yellow (C) spheres are cage *CPs* (color online).

**Figure 9**: Pressure variation of the order parameter *l* (the distance between the ring *CP* and the cage *CP*) (color online).

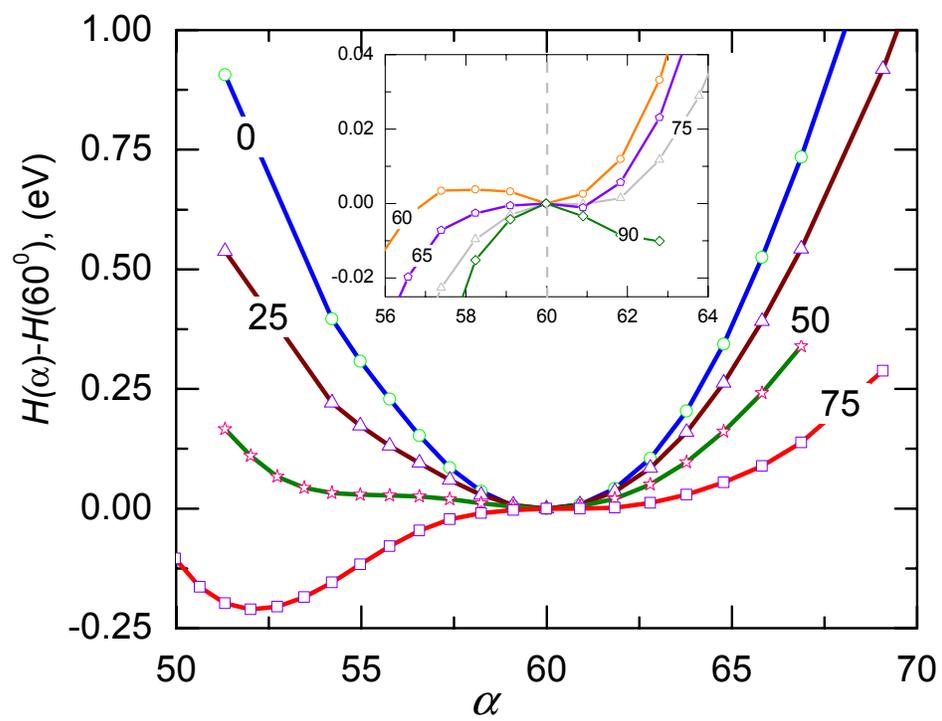

**Figure 1**

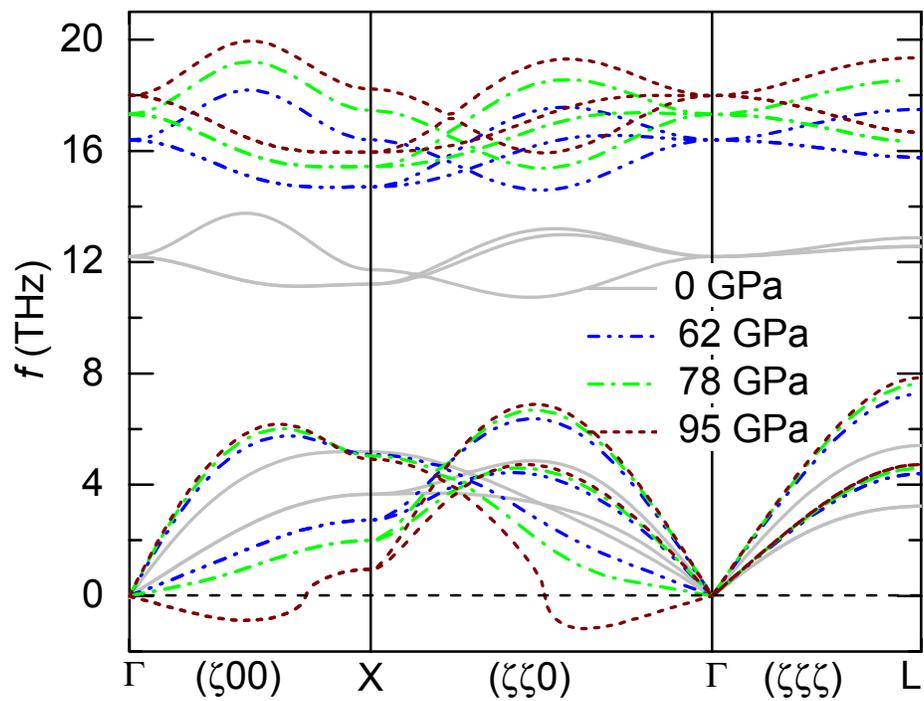

**Figure 2**

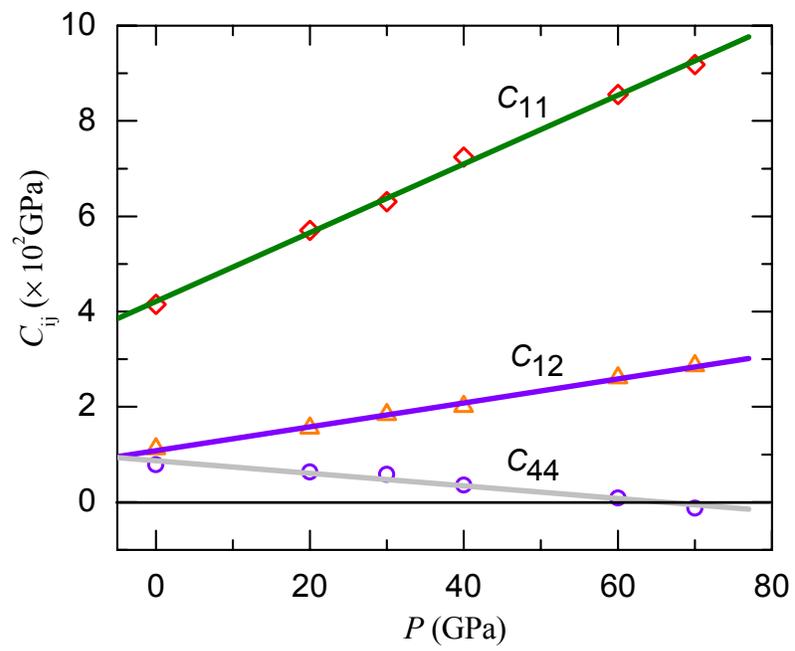

**Figure 3**

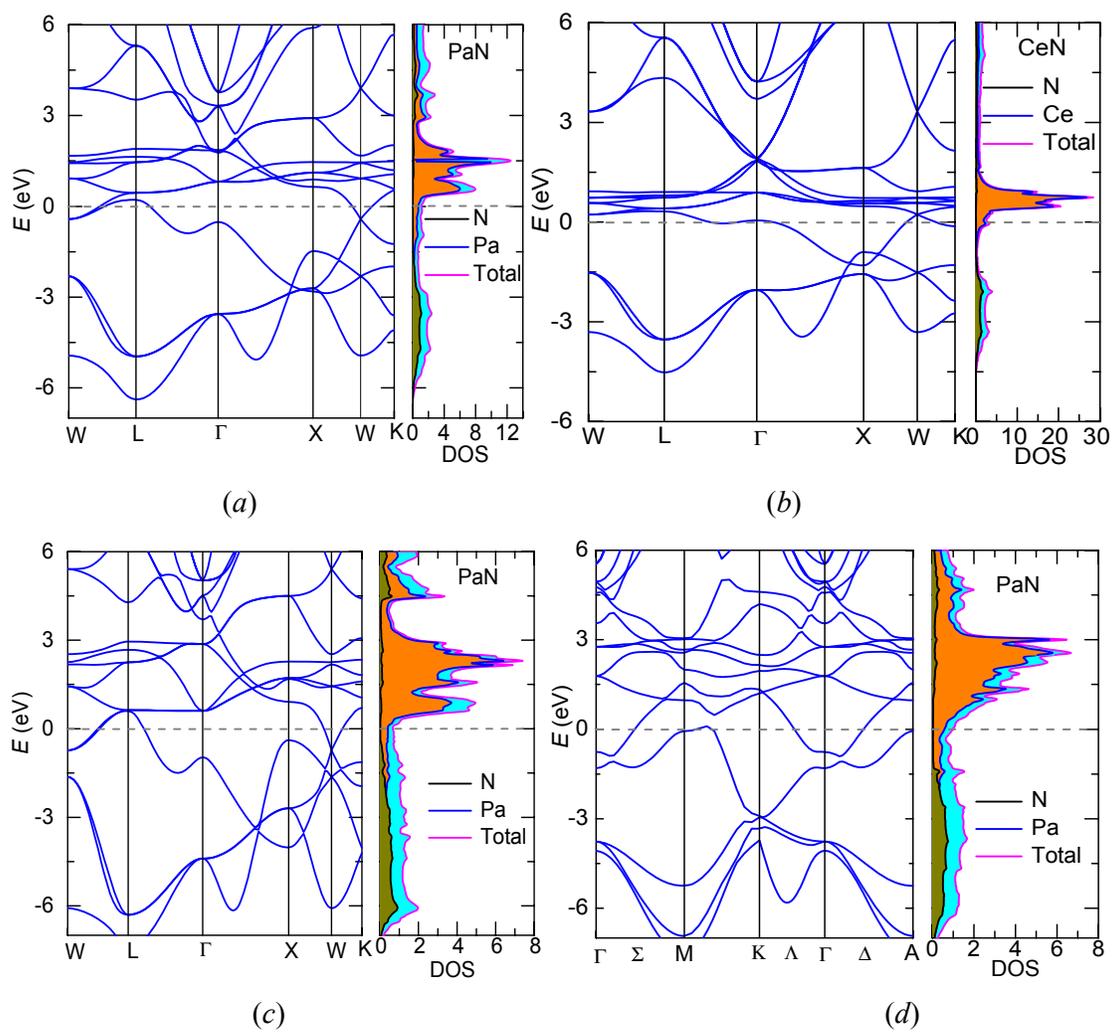

**Figure 4**

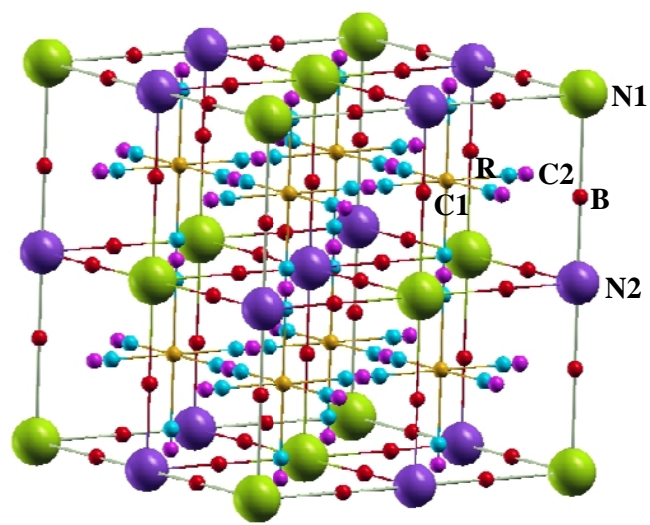

**Figure 5**

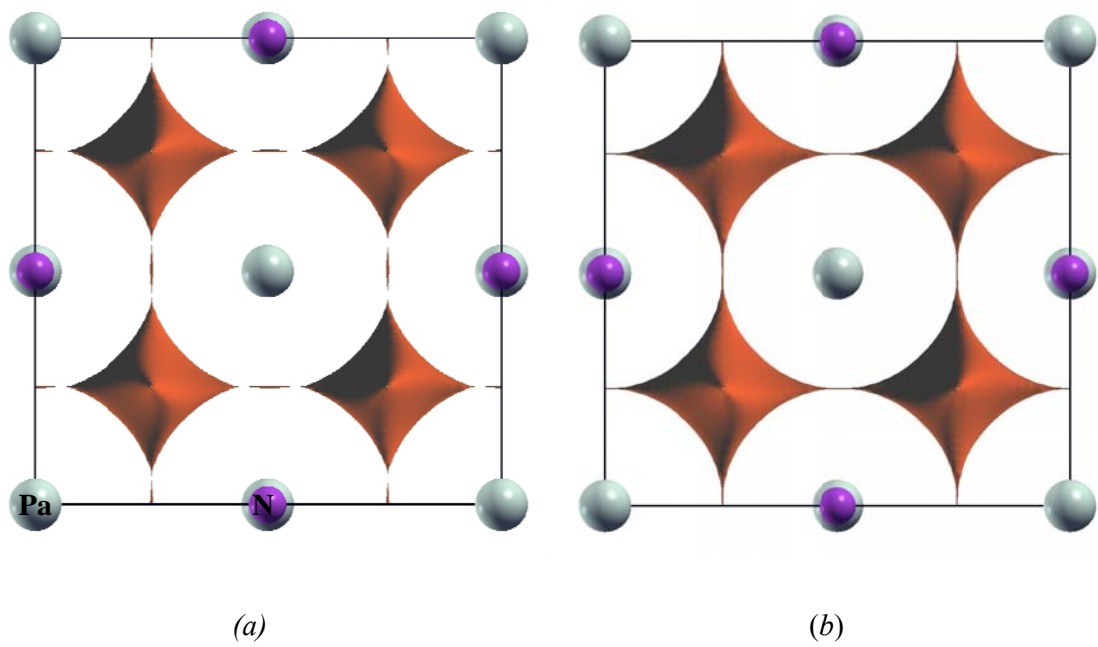

*(a)*          *(b)*

**Figure 6**

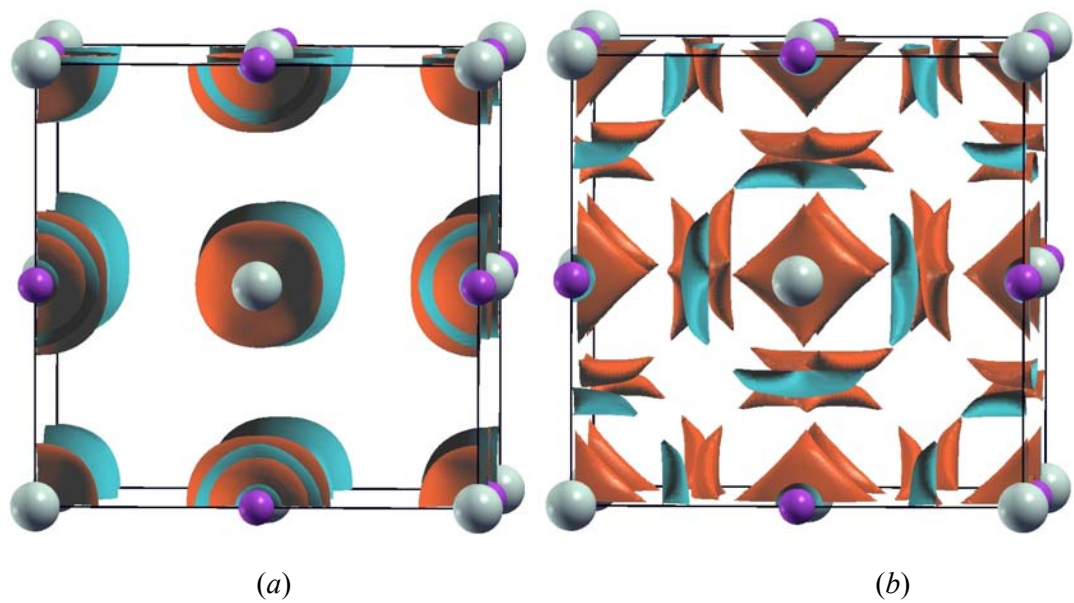

(*a*)　　　　　　　　　　　　　　(*b*)

**Figure 7**

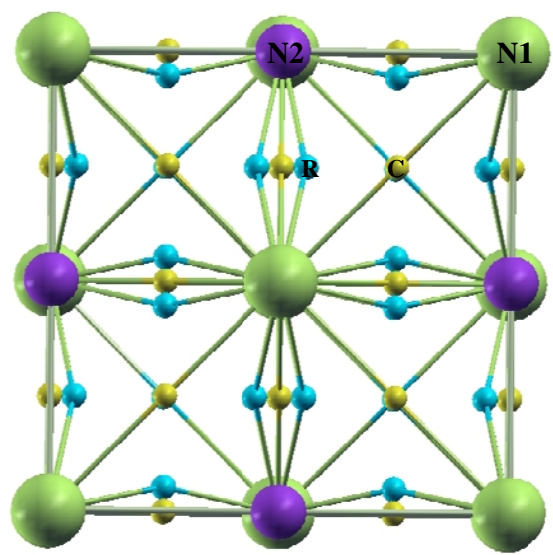

**Figure 8**

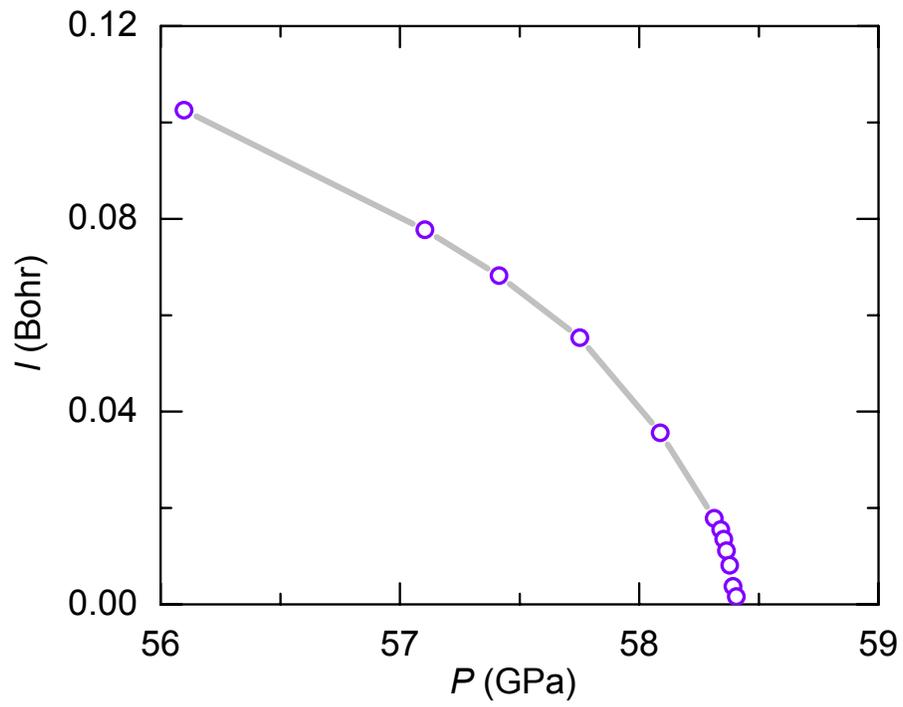

**Figure 9**